\documentclass[twocolumn,epjc3]{svjour3}
\smartqed  
\RequirePackage{graphicx}
\RequirePackage{color}
\RequirePackage{txfonts}
\UseRawInputEncoding

\begin{document}

\title{Constraining Yukawa-type interaction and coupling constant of axionlike
particles to nucleons from recent measurement of the Casimir-Polder interaction}

\titlerunning{Constraining Yukawa-type interaction and coupling constant of axionlike
particles}

\author{
G.~L.~Klimchitskaya\thanksref{addr1,addr2}\and
V.~M.~Mostepanenko\thanksref{addr1,addr2,e1}}

\authorrunning{G.~L.~Klimchitskaya and V.~M.~Mostepanenko}

\thankstext{e1}{e-mail: vmostepa@gmail.com (corresponding author)}

\institute{
Central Astronomical Observatory
at Pulkovo of the Russian Academy of Sciences,
St.Petersburg, 196140, Russia \label{addr1}
\and
 Peter the Great Saint Petersburg
Polytechnic University, Saint Petersburg, 195251, Russia
\label{addr2}
}

\date{Received: 28 February 2025  / Accepted: 22 April 2025}
%
\maketitle

\abstract{
We derive constraints on the parameters of the Yukawa-type
interaction and on the coupling constant of axionlike
particles to nucleons from the results of recent diffraction
experiment on measuring the Casimir-Polder interaction between
Ar atoms and a silicon nitride nanograting. It is shown that
within the interaction range from 1 to 2~nm the obtained
constraints are by up to a factor of 33.4
stronger than all the other ones found previously from measurements
of the Casimir force. The derived constraints are weaker in
strength than those deduced from the experiments on neutron
scattering. The constraints on the coupling constants of
axionlike particles to nucleons following from the diffraction
experiment are up to a factor of {24.2} stronger withing
the range of axion masses from 32.4 to 100~eV than the previously
derived ones from experiments on measuring the Casimir force.
They are weaker only in comparison to the constraints found
from the experiment using the beams of molecular hydrogen.
The potential of the Casimir effect for obtaining stronger
constraints on the parameters of hypothetical interactions
is discussed.}

\section{Introduction}

The interaction of Yukawa type between two point masses
arises due to an exchange of light scalar particles having
the mass $m$~\cite{1}. The existence of
such particles (dilatons, moduli, chameleons, etc.) is
predicted by the Kaluza-Klein theory, string theory, and
supergravity~\cite{2,3,4,5,6,7,8}. The Yuka\-wa-type
potential is also predicted {in the multi-dimensional
models}~\cite{9,10,11,12}.
The interaction range $\lambda$ of the Yukawa-type potential
{is caused by either the Compton wavelength of a scalar particle
$\lambda = \hbar /(mc)$ or by} the characteristic size of the compact
manifold {$\lambda \sim R_{\star}$.

The strongest} constraints on the parameters $\alpha$ and
$\lambda$ of the Yukawa-type potential were obtained from
different experiments. Thus, for $\lambda$ varying from
$2 \times 10^{-2}$~nm to 17.4~nm the strongest constraints
are placed by the experiment on neutron scattering~\cite{13},
from 17.4~nm to 40~nm -- by measurements of the effective
Casimir pressure using the micromechanical torsional
oscillator~\cite{15}, from 40~nm to 8~$\muup$m -- by the
differential force {measurements~\cite{16}}, from $8~\muup$m
to 1~cm -- by the Cavendish-type experiments~\cite{17,18,19,20,21},
and for $\lambda > 1$~cm -- by the E\"{o}tv\"{o}s-type
experiments~\cite{22,23}.

Axions are the pseudoscalar particles. Originally, they were
predicted in \cite{24,25}
{ as a consequence of violation of the
Peccei-Quinn symmetry, which was postulated~\cite{26} for a
solution of strong $CP$ problem in QCD by making the
$CP$-violating $\theta$-term sufficiently small}. In this
approach, axions are the Nambu-Goldstone bosons, which become
massive { because} the Peccei-Quinn symmetry
{ is explicitly broken}.
As a result, axions can be considered as an extension of the
Standard Model of elementary particles which awaits its
experimental confirmation.

The axion field interacts with the fermion fields by means
of the pseudovector coupling~\cite{27,28}. Along with axions,
the Grand Unified Theories introduce the axionlike particles,
which are not the Goldstone bosons and interact with fermions
via the preudoscalar coupling~\cite{27,28,29}. An exchange
of one axion or axionlike particle between two nucleons
results in the spin-dependent interaction
potential~\cite{30,31}, whereas an exchange of two axionlike
particles leads to the potential independent of the spins
of nucleons~\cite{32}.

For $m_a$ varying from $10^{-4}~\muup$eV to 1~$\muup$eV
the strongest constraints on $g$ are placed by measuring the
spin-dependent forces between neutrons by means of
magnetometer~\cite{33}, from 1~$\muup$eV to 0.676~meV -- by
the Cavendish-type experiments~\cite{19,20} measuring the
spin-independent forces, from 0.676~meV to 4.9~meV -- by
measuring the spin-independent forces of the gravitational
strength using the torsional oscillator~\cite{34,35,36},
from 4.9~meV to 0.5~eV -- by the differential measurements
of the spin-independent {forces~\cite{16,37},}
from 0.5~eV to 200~eV -- by measuring
the spin-dependent forces between protons in the beam of
molecular hydrogen~\cite{38,39}, and from 200~eV to 1000~eV --
by measuring these forces between nucleons in deuterated
molecular hydrogen~\cite{39}. In addition to the laboratory
experiments, strong constraints on the coupling constant of
axionlike particles to nucleons were obtained from the
astrophysical data~{\cite{28,40,40a}}. Numerous experiments
are also devoted to constraining the coupling constants of
axions and axionlike particles to electrons and photons
(see \cite{40a,41,42} for a review).

In this paper, we obtain constraints on the Yukawa-type
potential and on the coupling constant of axionlike
particles to nucleons, which are placed by the recent
diffraction
experiment on measuring the Casimir-Polder interaction
between Ar atoms and a silicon nitride {nanograting~\cite{43}.
It is shown} that the metrological accuracy reached in the
experiment \cite{43} makes it possible to strengthen
the constraints on the Yukawa-type interaction within the
interaction range from $\lambda = 1$~nm to
$\lambda = 2$~nm by up to a factor of 33.4
as compared to other constraints following from measurements
of the Casimir {force. It is also} shown that in the
region of masses of axionlike particles
$32.2~\mbox{eV} <m_a< 100~$eV
the experiment \cite{43} on measuring the
Casimir-Polder force leads to by up to a factor {24.2}
stronger constraints on the coupling constant of axionlike particles
to nucleons than those obtained previously from measurements
of the Casimir force.

The paper is organized as follows. In Sect.~2, we briefly
discuss the main features of the experiment \cite{43}
used for obtaining the constraints on hypothetical
interactions which may coexist with the Casimir-Polder
interaction. In Sect.~3, the constraints on the Yukawa-type
interaction are obtained. Section~4 is devoted to the
derivation of constraints on the coupling constant of
axionlike particles to nucleons. In Sect.~5, our conclusions
and a discussion are presented.

Below we use the system of units where $\hbar = c = 1$.

\section{Accurate measurement of the Casimir-Polder interaction in
diffraction experiment}
\newcommand{\ve}{\varepsilon}

In \cite{43}, the Casimir-Polder interaction potential was investigated
by means of diffraction of slow Ar atoms (with velocities below 16~m/s) on
a Si$_3$N$_4$ nanograting. An analysis of the measurement data of metrological
quality using the advanced statistical methods and a careful account of various
systematic effects were based on the previously developed description of atomic
diffraction on a nanograting \cite{49,50,51}.

It has been known that in the nonrelativistic limit (actually, at separations
$z\lesssim 3~$nm) the atom plate interaction potential is given by \cite{53}
\begin{equation}
V_{\rm C-P}(z)=-\frac{C_3}{z^3}, \quad
C_3=\frac{1}{4\pi}\int_0^{\infty}\!\!d\xi\alpha(i\xi)
\frac{\ve(i\xi)-1}{\ve(i\xi)+1},
\label{eq1}
\end{equation}
\noindent
where $C_3={\rm const}$, $\alpha(\omega)$ and $\ve(\omega)$ are the frequency-dependent
dynamic polarizability of an atom and dielectric permittivity of wall material
calculated along the imaginary frequency axis. At separations  $z\lesssim 1~$nm
there are corrections to the potential (\ref{eq1}) due to the discrete atomic structure of
the plate surface and the quantum effect of exchange repulsion described by the
Lennard-Jones potential  $V_{\rm L-J}=C_{\rm rep}/z^9$. Their role was included in
the balance of systematic errors.

At separations $z>3~$nm the relativistic effects come into play and the
Casimir-Polder interaction between an atom and a plate is given by the Lifshitz
formula \cite{53,54}, which can be represented in the form
\begin{equation}
V_{\rm C-P}(z)=-\frac{C_3f(z)}{z^3},
\label{eq2}
\end{equation}
\noindent
where
\begin{eqnarray}
&&
f(z)=\frac{1}{8\pi C_3}\int_0^{\infty}\!\!d\xi\alpha(i\xi)
\int_{2z\xi}^{\infty}\!\!dye^{-y}
\label{eq3} \\
&&~~~\times
\left[(y^2-2z^2\xi^2)r_{\rm TM}(i\xi,y)-2z^2\xi^2r_{\rm TE}(i\xi,y)
\right],
\nonumber
\end{eqnarray}
\noindent
and the reflection coefficients for the transverse magnetic (TM) and
transverse electric (TE) polarizations of the electromagnetic field
are given by
\begin{eqnarray}
&&
r_{\rm TM}(i\xi,y)=\frac{\ve(i\xi)y-\sqrt{y^2+4z^2\xi^2
[\ve(i\xi)-1]}}{\ve(i\xi)y+\sqrt{y^2+4z^2\xi^2
[\ve(i\xi)-1]}},
\nonumber \\
&&
r_{\rm TE}(i\xi,y)=\frac{y-\sqrt{y^2+4z^2\xi^2
[\ve(i\xi)-1]}}{y+\sqrt{y^2+4z^2\xi^2
[\ve(i\xi)-1]}}.
\label{eq4}
\end{eqnarray}

Note that for the gratings used in the experiment \cite{43} the diffraction
picture is formed at separations $z\lesssim 100~$nm from the grating faces.
Because of this, the thermal effects, which are not taken into account in
(\ref{eq2}) and (\ref{eq3}), are very small and can be neglected.
Computations show (see, for instance, \cite{55}) that the function
$f(z)$ defined in (\ref{eq3}) satisfies the condition
$f(z)\leqslant 1$.

The measurement results in \cite{43} were presented as the value of $C_3$
obtained from two measurement sets and its total error $\Delta C_3$ determined
at the 95\% confidence level. Besides the systematic errors mentioned above,
the error analysis took into account the systematic effects arising from
the finite area of the nanograting faces,
deviations of the nanograting profile from the rectangular geometry, surface
roughness, losses due to collisions of Ar atoms with the nanograting faces etc.
The resulting total systematic error was combined in quadrature with the random
error leading to
\begin{eqnarray}
&&
C_3={6.9\,\,}\mbox{meV\,nm}^3\pm\Delta C_3,
\nonumber \\
&&
\Delta C_3={1.2\,\,}\mbox{meV\,nm}^3,
\label{eq5}
\end{eqnarray}
\noindent
i.e., to {17.4\%} relative error. The measurement result (\ref{eq5}) is consistent
with computations by means of the Lifshitz formulas (\ref{eq1})--(\ref{eq3}) using
the optical data of Si$_3$N$_4$ which result in the values 7.37 \cite{57} and
7.42~meV\,nm$^3$ \cite{58}.

\section{Constraints on the Yukawa-type potential from measuring the Casimir-Polder
interaction in diffraction experiment}

The Yukawa-type interaction potential between the two pointlike masses $m_1$ and
$m_2$ is usually presented in relation to the Newtonian gravitational potential \cite{1}
\begin{equation}
V_{\rm Yu}(r)=\alpha e^{-r/\lambda}\,V_{\rm N}(r),
\qquad
 V_{\rm N}(r)=-\frac{Gm_1m_2}{r}.
 \label{eq6}
 \end{equation}
 \noindent
Here, $\alpha$ is the dimensionless strength of the Yukawa interaction, $r$ is the
distance between masses $m_1$ and $m_2$, and $G$ is the gravitational constant.

Now we calculate the Yukawa-type interaction energy $V_{\rm Yu}^{\rm AP}$ between an
Ar atom at the height $z$ and a nanograting face at $z=0$ approximated by the plate of
infinite large area and of $D\approx 100~$nm thickness. Integrating $V_{\rm YU}$ (\ref{eq6})
over the plate volume, one obtains \cite{53}
\begin{equation}
V_{\rm Yu}^{\rm AP}(z)=-2\pi Gm_{\rm A}\rho_{\rm P}\alpha\lambda^2
e^{-z/\lambda}\left(1-e^{-D/\lambda}\right).
\label{eq7}
\end{equation}
\noindent
Here, the mass of an Ar atom $m_{\rm A}=6.63\times 10^{-23}~$g and the
density of a silicon nitride nanograting is $\rho_{\rm  P}=3.17~\mbox{g/cm}^3$
\cite{56}.

The constraints on the parameters $\alpha$ and $\lambda$ of the Yukawa-type
interaction (\ref{eq7}) can be obtained from the fact that in the experiment
\cite{43} within the limits of the measurement error this interaction was not
observed, i.e.,
\begin{equation}
\left|\,V_{\rm Yu}^{\rm AP}(z)\,\right|<\frac{\Delta C_3\,f(z)}{z^3}.
\label{eq8}
\end{equation}

To be conservative, we replace $f(z)$ with its maximum value and obtain the
sought for constraints from the inequality
 \begin{equation}
\left|\,V_{\rm Yu}^{\rm AP}(z)\,\right|<\frac{\Delta C_3}{z^3}.
\label{eq9}
\end{equation}

\begin{figure}[t]
\vspace*{-2.5cm}
\centerline{\hspace*{2.cm}
\includegraphics[width=13.5cm]{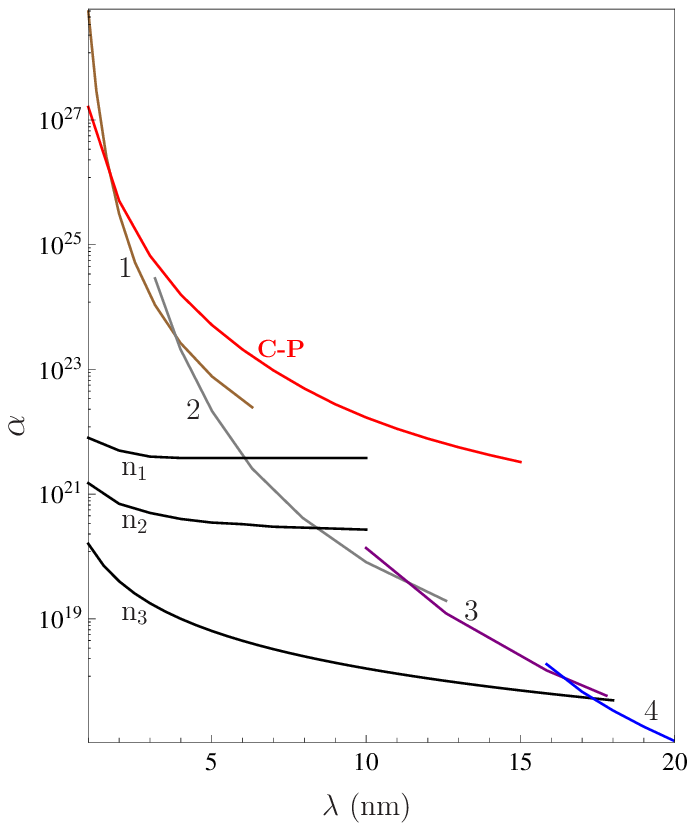}}
\vspace*{-7.9cm}
\caption{\label{fg1} Constraints on the interaction constant of the
Yukawa-type potential obtained from measurements of the
Casimir-Polder interaction in the diffraction experiment
are shown by the line labeled C-P versus the interaction
range. The previously obtained constraints from measuring
the Casimir force between a SiC plate and a borosilicate
microsphere, the lateral and normal Casimir forces between
corrugated surfaces, the effective Casimir pressure,
and from the experiments on neutron scattering are shown
by the lines 1, 2, 3, 4, and $n_1$, $n_2$, $n_3$, respectively.
The measurement data of each mentioned experiment exclude
the region of the $(\lambda,\,\alpha)$ plane above the
corresponding line and allow the region below it.}
\end{figure}
The resulting constraints are shown by the line labeled C-P in Fig.~\ref{fg1}
within the interaction range from $\lambda=1~$nm to $\lambda=15~$nm.
For each $\lambda$, the value of $z$ in (\ref{eq7}) and (\ref{eq9})
leading to the strongest constraint on $\alpha$ was chosen.
Here and below the region of the $(\lambda,\,\alpha)$ plane above the line
is excluded by the measurement data and below the line is allowed.

For comparison purposes, in Fig.~\ref{fg1} we also present the constraints
obtained from other measurements of the Casimir force and from experiments
on neutron scattering. Thus, the constraints obtained \cite{59} from the
experiment on measuring the Casimir force between a SiC plate and a
borosilicate microsphere \cite{60} are shown by the line 1.
The constraints of the line 2 were derived \cite{61} from measurements
of the lateral Casimir force between the sinusoidally corrugated surfaces
{\cite{63}.} In a similar way, the constraints of the line 3 were
derived \cite{64} from measurements of the normal to corrugated surfaces Casimir
force \cite{65,66}. Finally, by the line 4 we show the constraints following
from measurements of the effective Casimir pressure {\cite{15}}.
As to the lines labeled $n_1$, $n_2$, and $n_3$, they show the constraints
obtained from the experiments on measuring the scattered intensity of neutrons
on the xenon and helium gases \cite{47}, the gas of xenon atoms \cite{48},
and on the crystal of silicon \cite{13}, respectively.

As indicated in Fig.~\ref{fg1}, the novel constraints of the line C-P are
stronger than the constraints following from other measurements of the
Casimir force (lines 1, 2, and 3) over the interaction range from
$\lambda=1~$nm to 2~nm. Thus,  the largest strengthening, as compared
to the line 1, attained at 1~nm is by the factor of 33.4.

A comparison with the constraints following from the experiments on neutron
scattering shows that the constraints of the lines labeled
$n_1$, $n_2$, and $n_3$ remain stronger.

\section{Constraints on the coupling constant of axionlike particles to nucleons
from measuring the Casimir-Polder interaction in diffraction experiment}

{We} consider the spin-independent interaction potential between two nucleons
spaced $r$ apart, which is caused by the exchange of two axionlike particles \cite{32}
\begin{equation}
V_{a}(r)=-\frac{g^4}{32\pi^3 M^2}\,\frac{m_a}{r^2}\,K_1(2m_ar),
\label{eq10}
\end{equation}
\noindent
where $M=(m_p+m_n)/2$ is the mean nucleon mass and $K_1(x)$ is the modified Bessel function
of the second kind. Here and below, following \cite{31}, we assume that the coupling
constants of the axionlike particles to neutrons and protons are equal.

{The measurement of the Casimir-Polder interaction between
$^{87}$Rb atoms and a SiO$_2$ plate~\cite{44} has already
been used for constraining the axion-to-nucleon
interaction~\cite{45}. Later it was shown, however, that the
Casimir experiment \cite{15} result in stronger
constraints~\cite{46}.}

{Here,} we calculate an addition to the Casimir-Polder interaction potential
which arises due
to the exchange of two axionlike particles between the protons and neutrons belonging to
an Ar atom at the height $z$ and a nanograting face approximated by a plate in the same
way as in Sect.~3. For this purpose, we integrate expression (\ref{eq10}) in the polar
coordinates over the plate volume and obtain
\begin{eqnarray}
&&
V_{a}^{\rm AP}(z)=-\frac{m_a}{M^2M_{\rm H}}\left(\frac{g^2}{4\pi}\right)^2
\!C_{\rm A}C_{\rm P}
\nonumber\\
&&~~~~~~~~~~~~~~~~~~\times
\int_z^D\!\!\!\!dz_1\!\!\int_0^{\infty}\!\!\!\!\rho d\rho
\frac{K_1\left(2m_a\sqrt{\rho^2+z_1^2}\right)}{\rho^2+z_1^2}.
\label{eq11}
\end{eqnarray}
\noindent
Here the coefficients $C_{\rm A,P}$ for an Ar atom and for a Si$_3$N$_4$ plate are
defined as
\begin{equation}
C_{\rm A}=Z_{\rm A}+N_{\rm A}, \qquad
C_{\rm P}=\rho_{\rm P}\left(\frac{Z_{\rm P}}{\mu_{\rm P}}+
\frac{N_{\rm P}}{\mu_{\rm P}}\right),
\label{eq12}
\end{equation}
\noindent
$Z_{\rm A,P}$ and $N_{\rm A,P}$ are the numbers of protons and the mean numbers of
neutrons, respectively, in an Ar atom and a silicon nitride molecule. In doing so
\begin{equation}
Z_{\rm P}=3Z_{\rm Si}+4Z_{\rm N}, \qquad
N_{\rm P}=3N_{\rm Si}+4N_{\rm N},
\label{eq13}
\end{equation}
\noindent
where the numerical data for the quantities $Z_{\rm Ar}$, $Z_{\rm Si}$, $Z_{\rm N}$,
$N_{\rm Ar}$, $N_{\rm Si}$, $N_{\rm N}$ can be found in table 2.1 in \cite{1},
\begin{equation}
\mu_{\rm P}=3\mu_{\rm Si}+4\mu_{\rm N}, \qquad
\mu_{\rm Si}\equiv\frac{M_{\rm Si}}{M_{\rm H}},\qquad
\mu_{\rm N}\equiv\frac{M_{\rm N}}{M_{\rm H}},
\label{eq14}
\end{equation}
\noindent
$M_{\rm H}$ is the mass of atomic hydrogen,  $M_{\rm Si}$ and $M_{\rm N}$ are
the mean masses of Si and N atoms, and the density of silicon nitride
$\rho_{\rm P}$ was given in Sect.~3.

Substituting the values of all these parameters in (\ref{eq12}), one obtains
\begin{equation}
C_{\rm A}=39.985,\qquad
C_{\rm P}=3.196\times 10^3~\mbox{kg\,m}^{-3}.
\label{eq15}
\end{equation}

Introducing the new integration variable $t=\sqrt{\rho^2+z_1^2}$ instead of $\rho$
in (\ref{eq11}) and using the integral representation  for the modified Bessel
function \cite{67}
\begin{equation}
K_1(x)=x\int_1^{\infty}\!\!du\sqrt{u^2-1}\,e^{-ux},
\label{eq16}
\end{equation}
\noindent
we arrive at
\begin{eqnarray}
&&
V_{a}^{\rm AP}(z)=-\frac{2m_a^2}{M^2M_{\rm H}}\left(\frac{g^2}{4\pi}\right)^2
\!C_{\rm A}C_{\rm P}
\nonumber\\
&&~~~~~~~~~~~~~~~~\times
\int_z^D\!\!\!\!dz_1\!\!\int_{z_1}^{\infty}\!\!\!\!dt\!\!
\int_1^{\infty}\!\!du\sqrt{u^2-1}e^{-2m_atu}.
\label{eq17}
\end{eqnarray}
\noindent
Integrating here with respect to $t$ and then with respect to $z_1$, the final
result takes the form
\begin{eqnarray}
&&
V_{a}^{\rm AP}(z)=-\frac{1}{2M^2M_{\rm H}}\left(\frac{g^2}{4\pi}\right)^2
\!C_{\rm A}C_{\rm P}
\int_1^{\infty}\!\!du\frac{\sqrt{u^2-1}}{u}
\nonumber \\
&&~~~~~~~~~~~~~~~~~~~~~~~~\times
\left(e^{-2m_azu}-e^{-2m_aDu}\right).
\label{eq18}
\end{eqnarray}

Now, taking into account that the theoretical value for the Casimir-Polder
constant $C_3$ was confirmed by the measurement data within the limits of
total error $\Delta C_3$, the constraints on the parameters of axionlike
particles $g$ and $m_a$ can be obtained from the inequality
\begin{equation}
\left|\,V_{a}^{\rm AP}(z)\,\right|<\frac{\Delta C_3\,f(z)}{z^3}
<\frac{\Delta C_3}{z^3}.
\label{eq19}
\end{equation}

\begin{figure}[b]
\vspace*{-0.8cm}
\centerline{\hspace*{2.cm}
\includegraphics[width=14.5cm]{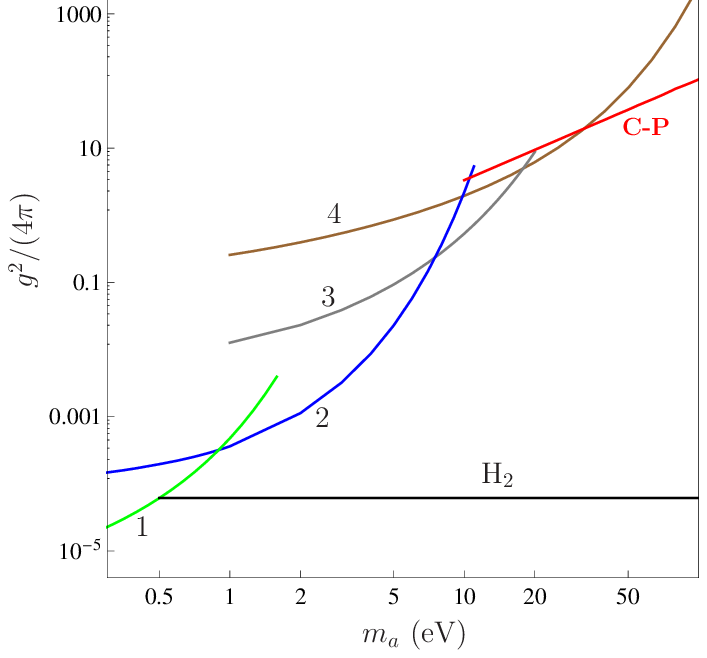}}
\vspace*{-11.7cm}
\caption{\label{fg2} Constraints on the coupling constant of axionlike
particles to nucleons obtained from measurements of the
Casimir-Polder interaction in the diffraction experiment
are shown by the line labeled C-P versus the axion mass.
The previously obtained constraints from the differential
force measurements, measuring the effective Casimir
pressure, the lateral Casimir force between corrugated
surfaces, the Casimir force between a SiC plate and a
borosilicate microsphere, and the spin-dependent forces
between protons in the beam of molecular hydrogen are
shown by the lines 1, 2, 3, 4, and H$_2$, respectively.
The measurement data of each mentioned experiment exclude
the region of the $(m_a,\,g^2/(4\pi))$ plane above the
corresponding line and allow the region below it.
}
\end{figure}
Similar to the case of Yukawa-type interaction, for each value of $m_a$, the value
of $z$ in (\ref{eq18}) and (\ref{eq19}) leading to the strongest constraint
on $g$ was chosen. The obtained constraints on the quantity $g^2/(4\pi)$ are shown
in Fig.~\ref{fg2} by the line labeled C-P.
Note that according to recent results \cite{67a}, under certain assumptions, the
constraints of this line are approximately valid not for only axionlike particles,
but also for massive QCD axions whose interaction with nucleons is described
by the pseudovector coupling.
In the same figure, the constraints
derived \cite{37} from the differential force measurements, where the contribution
of the Casimir force was nullified \cite{16}, are shown by the line 1.
By the line 2 in Fig.~\ref{fg2} the constraints are shown found \cite{46} from
measurements of the effective Casimir pressure using the micromechanical
oscillator {\cite{15}}. The constraints of the line 3 follow \cite{68}
from measurements of the lateral Casimir force between the sinusoidally corrugated
surfaces {\cite{63}}. Finally, the line 4 shows the constraints derived \cite{59}
from the experiment on measuring the Casimir force between a SiC plate and a
borosilicate microsphere. All these constraints obtained from measuring  the
Casimir force between the unpolarized bodies were obtained using the process of
the exchange of two axionlike particles between nucleons. As to the line in
Fig.~\ref{fg2} labeled H$_2$, it was derived \cite{39} from measurements of the
spin-dependent forces between  protons in the beam of molecular hydrogen \cite{38}.
These constraints exploit the exchange of one axion  or axionlike particle and,
as a result, are much stronger. {Note also that the possibility of
constraining the coupling of axions to both electrons and nuclei
in the $10^{-4}$~eV mass range using atomic transitions was proposed
in~\cite{69}.}

Although the constraints shown by the line C-P in Fig.~\ref{fg2} cannot compete with
those shown by the line labeled H$_2$,  within the range of masses of axionlike
particles from 32.2~eV to 100~eV they are stronger than the constraints of line 4
obtained previously from measurements of the Casimir force. The major strengthening
by the factor of {24.2} is reached for the axionlike particles of 100~eV mass.

\section{Conclusions and discussion}

In the foregoing, we have considered the constraints on the
Yukawa-type interaction predicted in many extensions of the
Standard Model and on the coupling constant of axionlike
particles to nucleons. Both these subjects are of considerable
interest because, by constraining the Yukawa-type interaction,
we obtain a valuable information about the proposed
multi-dimensional models, whereas the stronger constraints
on axionlike particles may help to clarify their status as
the most probable constituents of dark matter.

After a brief listing of the strongest constraints obtained
so far on the strength of Yukawa-type potential and the
coupling constant of axionlike particles to nucleons
 from the laboratory experiments, we discussed the
recent diffraction experiment~\cite{43} on measuring the
Casimir-Polder interaction between Ar atoms and a silicon
nitride nanograting. It was emphasized that this experiment
has many points in its favor by using the slow atomic beam
of noble argon gas and the metrological study of all possible
sources of both systematic and random errors.

According to our results, the measure of agreement between
the prediction of the Lifshitz theory for the Casimir-Polder
coefficient $C_3$ and its experimental value makes it
possible to strengthen the constraints on the
interaction constant of the Yukawa-type interaction within
the interaction range from 1 to 2~nm, as compared to
all previous measurements of the Casimir force. The major
strengthening by the factor of 33.4 holds at
$\lambda = 1$~nm.

We have also used the same measure of agreement between
experiment and theory for constraining the coupling
constant of axionlike particles to nucleons. It was shown
that in the range of masses of axionlike particles from
32.2 to 100~eV the obtained constraints are stronger by
up to a factor of {24.2} than those found previously from
measurements of the Casimir force. In this case, however,
the constraints following from the diffraction experiment
on measuring the Casimir-Polder interaction cannot compete
with the constraints found using the
exchange of one axion between protons in the beam of
molecular hydrogen  \cite{39}.

It should be underlined that by decreasing the contribution
of systematic errors to the total error in the diffraction
experiment (currently they contribute 16.2\% to the total
relative error of 17.2\% leaving only 1\% for the random
errors~\cite{43}) it will be possible to strengthen the
constraints on both the Yukawa-type interaction and the
coupling constant of axionlike particles to nucleons
following from this experiment. That is why measurements
of the Casimir force preserve their potential for
obtaining stronger constraints on the parameters of
hypothetical interactions.

\begin{acknowledgement}
This work was supported by the State Assignment for Basic Research
(project FSEG-2023-0016).
\end{acknowledgement}

\end{document}